\documentclass[aps,prb,twocolumn,groupedaddress,floatfix,showpacs]{revtex4}
\usepackage{graphics}
\usepackage{subfigure}
\usepackage{epsfig}
\usepackage{dcolumn}
\usepackage{bm}
\usepackage{overpic} 

\begin{document}

\title{Oxidation effects on graded porous silicon anti-reflection coatings}
\author{Annett Th\o gersen$^{1,2}$}
\author{Josefine H. Selj$^1$}
\author{Erik S. Marstein$^1$}

\address{$^1$Institute for Energy Technology, Department of Solar Energy, Instituttveien 18, 2008 Kjeller, Norway}
\address{$^2$SINTEF Materials and Chemistry, P.O.Box 124 Blindern, 0314 Oslo, Norway}

\date{\today}

\begin{abstract}

Efficient anti-reflection coatings (ARC) improve the light collection and thereby increase the current output of solar cells. By simple electrochemical etching of the Si wafer, porous silicon (PS) layers with excellent broadband anti-reflection properties can be fabricated. In this work, ageing of graded PS has been studied using Spectroscopic Ellipsometry, Transmission Electron Microscopy and X-ray Photoelectron Spectroscopy. During oxidation of PS elements such as pure Si (Si$^0$), Si$_2$O (Si$^+$), SiO (Si$^{2+}$), Si$_2$O$_3$ (Si$^{3+}$), and SiO$_2$ (Si$^{4+}$) are present. In addition both hydrogen and carbon is introduced to the PS in the form of Si$_3$SiH and CO. The oxide grows almost linearly with time when exposed to oxygen, from an average thickness of 0 - 3.8 nm for the surface PS. The oxidation is then correlated to the optical stability of multi-layered PS ARCs. It is found that even after extensive oxidation, the changes in the optical properties of the PS structures are small.

\end{abstract}

\maketitle

\section{Introduction}

Good anti-reflection coatings (ARC) can improve the light collection and thereby increase the efficiency of solar cells considerably. Porous Silicon (PS) multilayers have excellent broadband anti-reflective properties, and can be made using electrochemical etching of Si wafers in an electrolyte containing hydrofluoric acid (HF). However, the structure is very sensitive to etching parameters such as current density, electrolyte composition, temperature and substrate doping \cite{Zhang:ny, Rossow:ny}. 

Several groups have reported results from using PS as antireflection coating in solar cells \cite{Yuan:ny, Kwon:ny, Lipinski:ny, Yerokhov:ny, Wettling:ny, selj:2}. We have previously shown that graded PS ARCs with an effective reflection of 3 \% over the solar spectrum can be produced in p$^+$ material. PS has a very high internal surface area, 200-600 m$^2$/cm$^3$, and is therefore significantly more reactive than bulk Si\cite{Herino:PS}. In order for these structures to be used in a solar cell device, the electrical and optical properties must not degrade with ageing. Extensive oxidation has been shown by Canham et al. \cite{Canham:1}  to occur even in air ambient at room temperature, but strong dependence of the oxidation process on the porosity, morphology, thickness, and storage of the PS structures are observed. 

Oxidation of PS can take place both through continued SiO$_2$ formation and formation of O$_x$-Si-H groups. Grosman and Ortega \cite{Grosman:1} reported that the natural oxidation of PS forms SiO$_2$ together with Si-OH and SiO$_2$-OH, where Si is bounded to one or three oxygen atoms \cite{Grosman:1}, in addition to SiH$_x$ (x= 1, 2, 3) groups \cite{grosman:16, Rao:19}. Domashevskaya at al. \cite{Domashevskaya:13} found the presence of (CH$_3$CH$_2$)$_3$SiOH compounds in addition to amorphous Si (Si:H) in PS layers using XPS. More common compounds found in oxidized Si samples are Si$_2$O, SiO, Si$_2$O$_3$, and SiO$_2$ \cite{Peden:1, Himpsel:2, Thøgersen:3}. In this paper, Spectroscopic Ellipsometry (SE), Transmission Electron Microscopy (TEM) and X-ray Photoelectron Spectroscopy (XPS) have been used to study the various oxidation states present before and after ageing of PS, and the effect upon the optical properties.

\section{Experimental}

The electrochemical etching was performed using a double cell anodic etching system PSB Plus 4 from advanced micromachining tools (AMMT). Prior to etching, the wafers were dipped in 5\% HF to remove native oxide and to clean the samples. The electrolyte consisted of 49 wt \% HF and ethanol (C$_2$H$_5$OH) in a volume ratio 2:3, giving a HF concentration of about 20\%. The wafers used were boron doped, 300-350 $\mu$m thick, one side polished, monocrystalline Si with a (100) orientation. PS is formed on the polished side of the wafer without any additional texturing. The resistivity of the wafers was determined by four point probe to be 0.012-0.018 $\Omega$cm, which corresponds to a doping level of approximately 5x10$^{18}$. In order to simplify the identification and quantification of oxidation effects from the SE measurements, homogeneous PS layers were used. Five homogeneous PS samples were etched under galvanostatic conditions, at a current density of 50 mA/cm$^2$ for 15 s. The single layer samples have approximately the same average porosity as the graded PS ARCs. Etching of the graded PS ARC structures is also performed in galvanostatic mode, by stepwise variation of the current density. Details of the procedure, such as the duration and current density of each step are described elsewhere \cite{selj:2}. Only the current density during etching is different for the homogeneous and multilayered samples and the average porosities of the two are similar. 

The first homogeneous sample (Day 0) was rinsed in water, dried in N$_2$ and stored in aluminium foil during transportation to the XPS. This was done in order to achieve a minimally oxidized starting sample. All remaining samples were rinsed in ethanol and air dried. Ethanol has a reduced surface tension compared to water and therefore reduces the risk of surface cracking and flaking of the PS films. The samples were stored in in air at room temperature.

Depth profile XPS was carried out by sputtering the surface using an Ar beam. Since this process alters the surface of the sample, one single sample could not be used for all measurements. Five nominally identical samples were therefore used for XPS. We have previously shown that the PS layers used in this work are reproducible to within an uncertainty of 1.5 \% (90 \% cl.) in thickness and porosity \cite{Selj:1}.

The different elements present after the oxidation of PS have been studied by XPS. XPS was performed in a KRATOS AXIS ULTRA$_{DLD}$ using monochromated Al K$\alpha$ radiation (h$\nu$ =1486.6 eV) on plane-view samples at zero angle of emission (vertical emission). The x-ray source was operated at 1 mA and 15 kV. Depth profile sputtering was performed using a 4 kV ion gun, with a current of 100 $\mu$A and a 500s cycle time. The etch depth was estimated to be approximately 100 nm. The mean free path ($\lambda$) of Si-2p electrons in Si is 3.18 nm. This means that the photoelectron escape depth in Si is 3$\lambda \cos(\Theta$) = 9.54 nm, with $\Theta = 0$. The spectra were peak fitted using CasaXPS \cite{casa:xps} after subtraction of a Shirley type background. Cross-sectional TEM samples were prepared by ion-milling using a Gatan precision ion polishing system with 5 kV gun voltage. The samples were analysed by HRTEM in a 200 keV JEOL 2010F microscope with a Gatan imaging filter and detector. 

Ellipsometry measurements were carried out using a Woollam variable angle spectroscopic ellipsometer (WVASE) in the wavelength range 300$-$1200 nm and at the angles of incidence 60$^\circ$, 65$^\circ$, 70$^\circ$, 75$^\circ$, and 80$^\circ$. Depolarization data is also collected for all samples. A model consisting of Si and voids in a Bruggeman effective medium approximation (EMA) provided a good fit to all the ellipsometric data sets \cite{Pettersson:JHS, Rossow:JHS}. Finally, reflectance was measured using a Standford Research System set-up in the wavelength range 400$-$1100 nm.

\section{Results and Discussion}

\begin{figure}
  \begin{center}
    \includegraphics[width=0.4\textwidth]{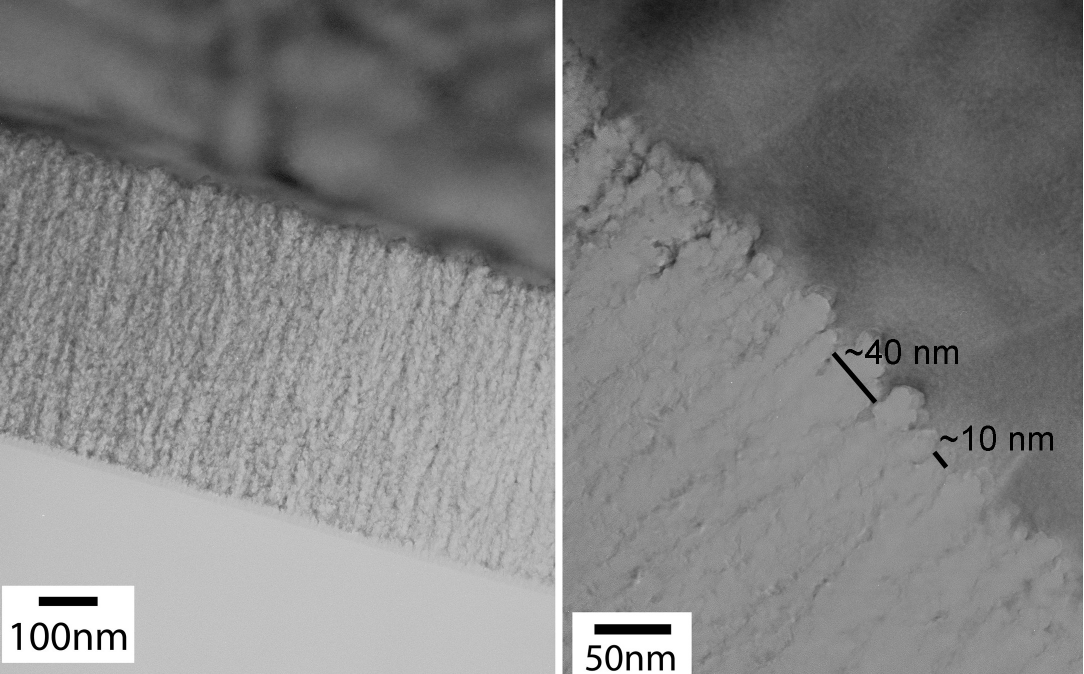}
   \caption{TEM image of the porous silicon layer. The figure on the left presents an overview of the PS layer, and the one on the right shows the etch craters and PS walls.}
    \label{figure:tem}
  \end{center}
\end{figure}

\begin{figure}
  \begin{center}
    \includegraphics[width=0.5\textwidth]{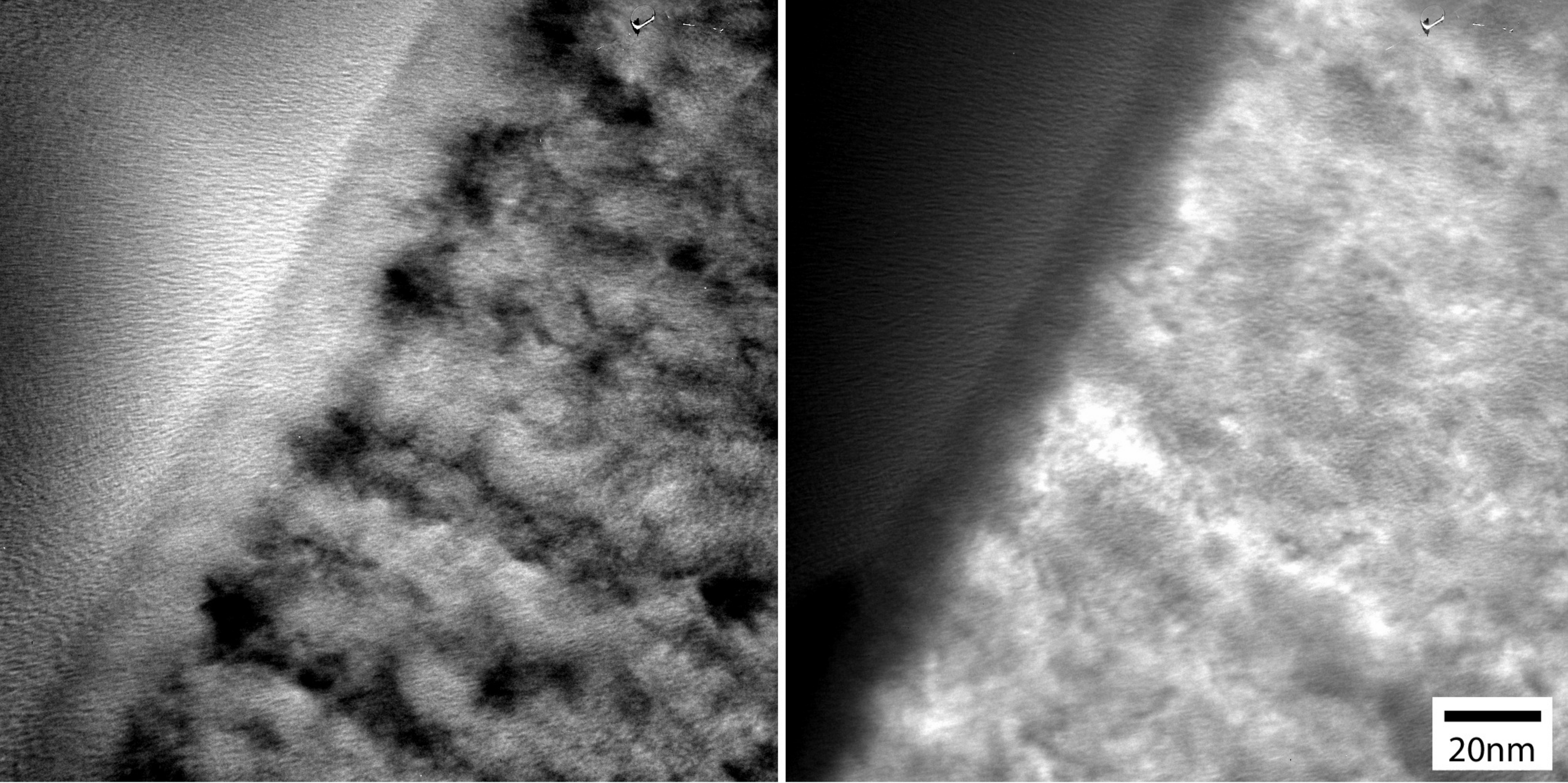}
   \caption{Energy Filtered TEM images showing A) filtered image of the plasmon peak of SiO$_2$ at 23 eV, and B) the plasmon peak of Si at 16 eV.}
    \label{figure:2}
  \end{center}
\end{figure}

Two representative TEM images of the PS samples are shown in Figure \ref{figure:tem}. The PS layers are about 500 nm thick, with a pore size between 10-40 nm in diameter. A detailed study of the oxidation states and compounds present during ageing has been carried out using XPS. The SiO$_x$ thickness has also been calculated, and correlated to changes in porosity and refractive index during ageing. Energy Filtered TEM images in Figure \ref{figure:2} shows the SiO$_x$ thickness of the sample aged 42 days. Figure \ref{figure:2}a filters the plasmon peak of SiO$_2$ at 23 eV, while Figure \ref{figure:2}b the plasmon peak of Si at 16 eV. The images clearly shows that the 15 nm thick layer outside the PS is due to SiO$_x$ and not to amorphous.  

\subsection{Peak-fitting Porous Silicon spectra}

\begin{figure}
  \begin{center}
    \includegraphics[width=0.5\textwidth]{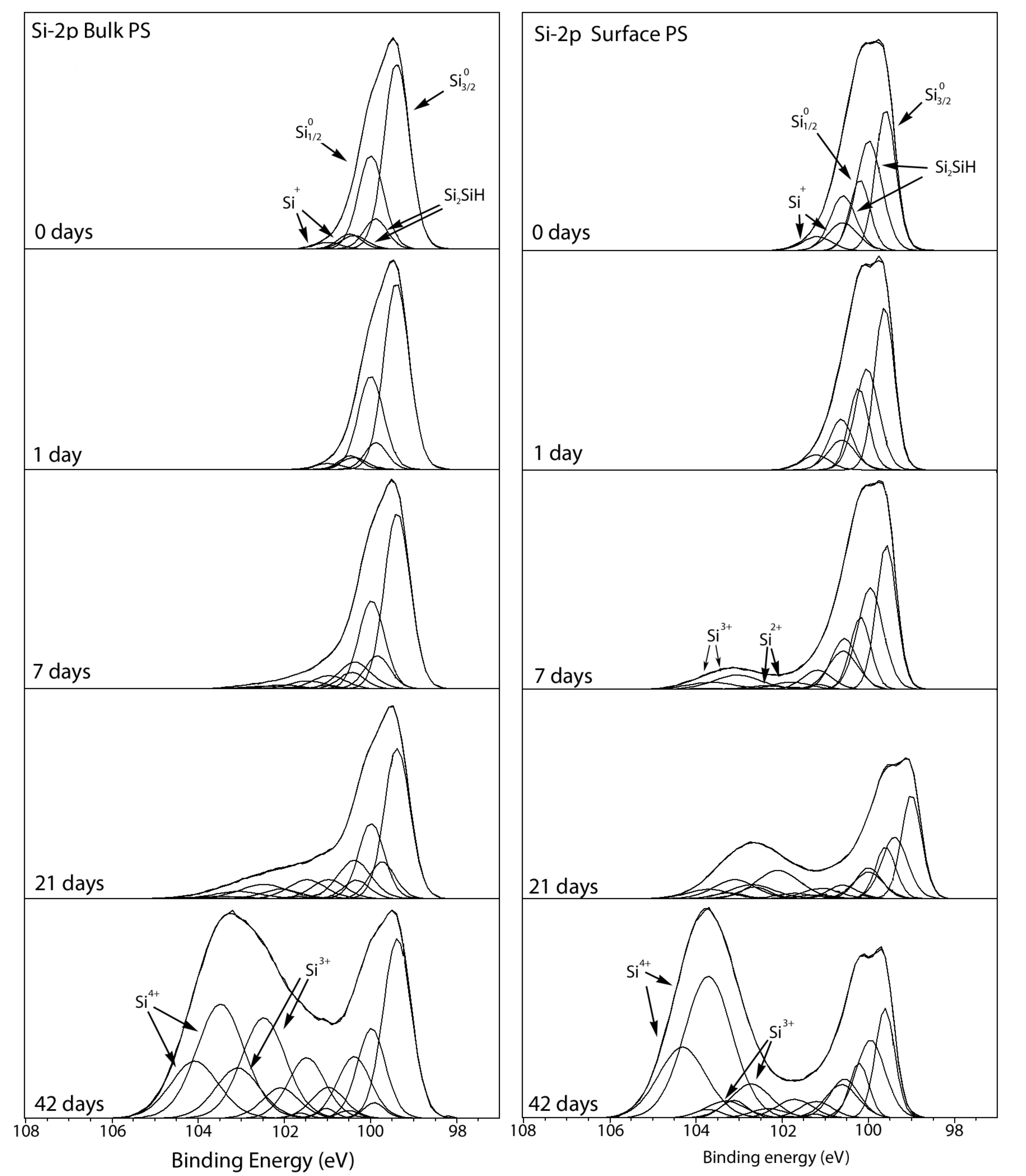}
   \caption{XPS depth profile spectra of the Si-2p peaks of A) bulk PS, and B) surface PS. The spectra are normalized in order to visualize all peaks.}
    \label{figure:3}
  \end{center}
\end{figure}

The Si-2p XPS spectra of both bulk and surface PS are shown in Figure \ref{figure:3}, for samples exposed to oxygen for 0, 1, 7, 21 and 42 days. The XPS measurements of surface PS will collect photoelectrons down to a depth of about 20 nm, while the XPS spectra of bulk PS have been obtained after sputtering down to a depth of about 100 $\pm$ 20 nm. In order to find all possible compounds present in the PS samples, a detailed peak fitting was carried out. 

Cerofolini et al. \cite{Cerofolini:6} and Lu et al. \cite{Lu:18} showed a detailed peak fitting of the Si-2p spectra. A pure Gaussian function was shown to have the best fit for all oxide peaks. However, for bulk Si, an asymmetrical peak with a tail length of 6.5, tail scale of 0.6\% and 70\% Gaussian has been found to provide the best fit (GL(30)T(6.5)). The Si-2p$_{1/2}$ and Si-2p$_{3/2}$ were first fitted using this data, with a relative intensity of 1:2 (Si-2p$_{1/2}$ / Si-2p$_{3/2}$) \cite{Eschner:17}. The SiO$_x$ peaks (Si$^+$, Si$^{2+}$, Si$^{3+}$ and Si$^{4+}$) were then fitted with a pure Gaussian function GL(0). The full width half maximum (FWHM) was set at 0.65 eV as has been reported for the Si-2p$_{1/2}$ and Si-2p$_{3/2}$ by Peden et al.\cite{Peden:1} and Himpsel et al. \cite{Himpsel:2}. The binding energy of the 2p$_{1/2}$ and 2p$_{3/2}$ has been reported to be between 99.0 $-$ 100.5 eV \cite{Cerofolini:6}, while the binding energy of Si-2p for SiO$_2$ is 103.6 eV \cite{Wagner:5}. The chemical shift (EB(Si$^{4+}$) - EB(Si$^0$)) between the Si-2p$_{1/2}$ and Si-2p$_{3/2}$ is reported to be 0.6 eV with a FWHM of 0.65 eV \cite{Peden:1, Himpsel:2}. The FWHM of the other oxidation states of Si were then found by comparing their relative sizes to our previous study for non-monochromatic XPS on Si \cite{Thøgersen:3}. The appropriated FWHM values for the Si$_2$O, SiO, Si$_2$O$_3$, and SiO$_2$ were found to be 0.8 eV, 1.1 eV, 1.1 eV, and 1.2 - 1.5 eV respectively. There is a small difference in FWHM for surface PS and bulk PS, 0.55 eV and 0.70 eV respectively. The XPS spectra of surface and bulk PS were fitted to Si-2p$_{3/2}$ peaks with a binding energy of 99.4, 100.4 eV, 101.4 eV, 102.5 eV, 103.6 eV, and 99.7 eV. The five first peaks correspond to pure Si (Si$^0$), Si$_2$O (Si$^+$), SiO (Si$^{2+}$), Si$_2$O$_3$ (Si$^{3+}$), and SiO$_2$ (Si$^{4+}$). That leaves the peak with a binding energy of 99.7 eV.

\subsubsection{Discussion of peak positions and FWHM}

The fitted peak positions and FWHM is presented in Table \ref{table:eb} for all oxidation states. Oxidation states such as Si$_3$SiH, Si$_2$SiH$_2$ or Si$_3$SiC may also occur due to HF etching and C contamination at the surface \cite{Cerofolini:6}. Si$_3$SiH can be due to elemental silicon bonded to one hydrogen atom, and has 0.3 eV higher binding energy than for elemental silicon, while the compound Si$_2$SiH$_2$ has a 0.57 eV higher binding energy \cite{Cerofolini:6}. The peak located at 99.7 eV has a 0.3 eV higher binding energy than elemental Si, which means that it is probably due to Si$_3$SiH. 

\begin{table}[h!]
 \caption{The binding energy (E$_B$) and full width half maximum (FWHM) of the oxidation states present in the samples, The EB$_{_{3/2}}$ and EB$_{_{1/2}}$ is given for the Si oxidation states.}	
\begin{tabular}{llll}
\hline
Ox. state & EB$_{_{3/2}}$ & EB$_{_{1/2}}$ & FWHM \\
 & [eV] & [eV] & [eV] \\
 \hline
Si2p: Si$^0$ & 99.4 & 100.0 & 0.55-0.7 \\
Si2p: Si$^{+}$ & 100.4 & 101.0 & 0.8 \\
Si2p: Si$^{2+}$ & 101.4 & 102.0 & 1.1 \\
Si2p: Si$^{3+}$ & 102.5 & 103.1 & 1.1 \\
Si2p: Si$^{4+}$ & 103.6 & 104.2 & 1.2-1.5 \\
Si2p: Si$_3$SiH & 99.7 & 100.3 & 0.7 \\
\hline
O1s:SiO$_x$ & $\sim$532.5 & & 1.5 \\
O1s: C-OH & $\sim$ 533.4 & & 1.4 \\
C-O-C & & & \\
O1s:C=O & $\sim$ 531.8 & & 1.4 \\ 
\hline
C1s: C & 284.5 & & 1.2 \\
C1s: C-O-C & 286.5 & & 1.3 \\
\hline 
\end{tabular}
\label{table:eb}
\end{table}

\begin{figure}  
	\begin{center}
    \includegraphics[width=0.5\textwidth]{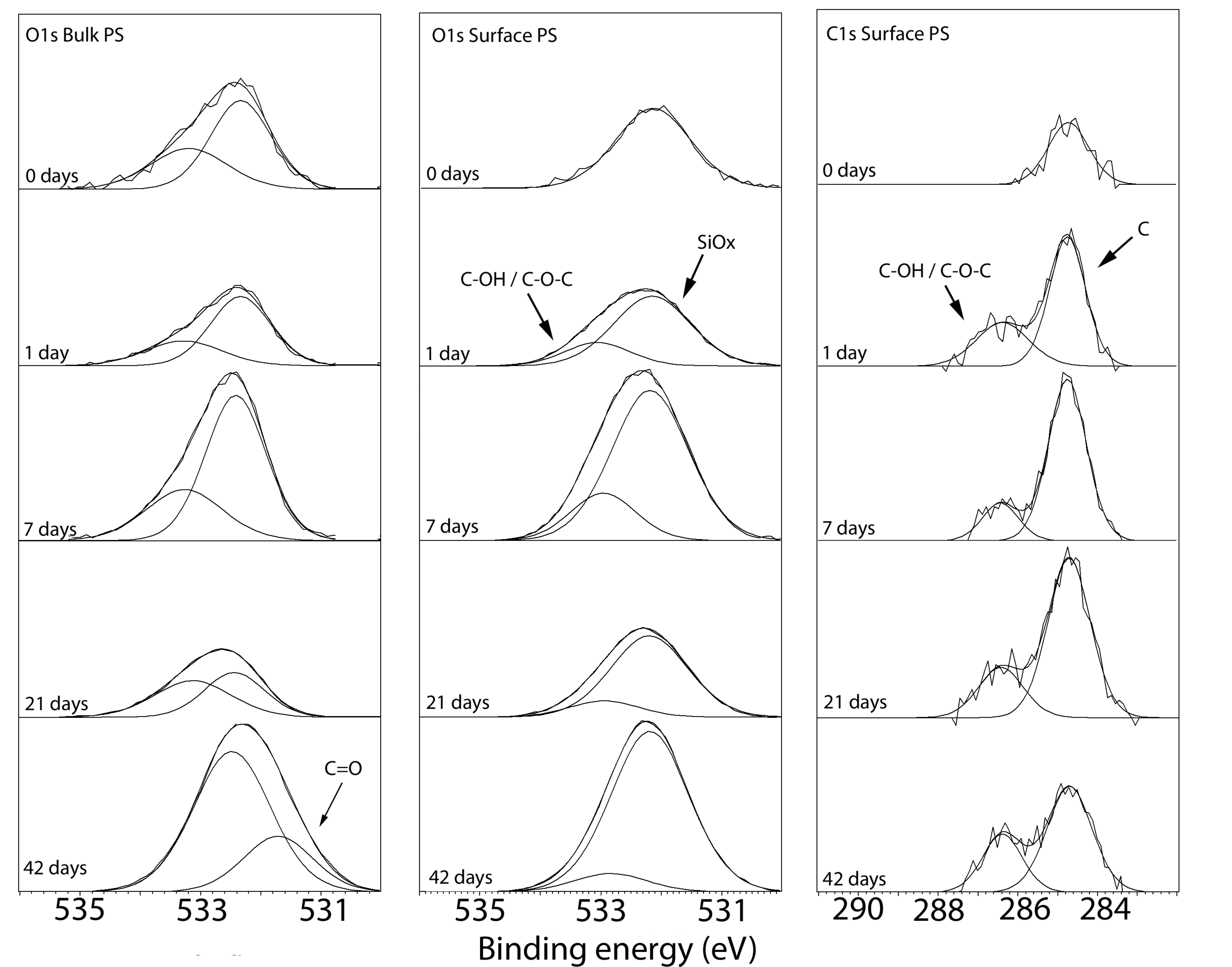}
   \caption{XPS depth profile spectra of A) the O1s peak of bulk PS, B) the O1s peak of surface PS, and C) the C1s peak of surface PS. The spectra are normalized in order to visualize all peaks.}
    \label{figure:4}
  \end{center}
\end{figure}

XPS spectra of the O1s peak for surface and bulk PS is presented in Figure \ref{figure:4} together with the surface PS spectra of the C1s peak. The O1s spectrum of both surface and bulk PS have two clear peaks. The O1s peak from SiO$_2$ has a reported binding energy of 533.05 eV \cite{Hollinger:SiO2}. The O1s binding energy from C=O is reported to be between 531.2 $-$ 531.6 eV, from C-OH and/or C-O-C between 532.2 $-$ 533.4 eV, and for chemisorbed oxygen and perhaps some adsorbed water between 534.6$-$535.4 eV \cite{yue:xps}. The large peak at a binding energy of 532.5 eV seems to fit well with SiO$_x$. The smaller peak at higher binding energy may therefore be due to C-OH and/or C-O-C. Bulk PS after 42 days shows an additional peak at 531.7 eV, which may be due to C=O. Figure \ref{figure:4} also shows the C1s peak of surface PS. The spectra can be fitted with two peaks located at a binding energy of 284.5 eV and 286.5 eV. Graphitic carbon has been reported to have a C1s binding energy of 284.6 eV, while CO between 290.4 $-$ 290.8 eV \cite{yue:xps}. From the O1s peaks the composition C-OH and/or C-O-C was identified. This composition can have a slightly smaller binding energy than CO, which would fit better with the observed binding energy. The two peaks may therefore correspond to pure C and C-OH and/or C-O-C.

There was a 0.15 eV difference in FWHM of Si$^0$ between surface and bulk. Guerrero-Lemus et al \cite{GuerreroLemus:7} have studied the composition of PS using XPS and FTIR. They reported that XPS peaks in the Si-2p spectra can be attributed to Si-H bonds, in addition to amorphization of Si, due to a broadening of the Si peak. The presented Si-H peak is not shown clearly in the XPS spectrum, especially since there has been no peak fitting. In addition, no binding energy or FWHM values of the fitted peaks have been presented and compared to reference values. Ley et al. \cite{Ley:8} studied amorphous Si and found that the broadening in the XPS Si-2p peaks was due to a distribution of chemically shifted Si-2p lines. This shift occurred because of random charge fluctuations as the result of bond length variations in the amorphous network. However, a decrease of broadening was found in proportion to the number of hydrogen atoms attached to the silicon atom (Si, Si-H, Si-H$_2$, Si-H$_3$) due to charge transfer from Si to H. A small difference in FWHM between pure Si and Si-H was found to be 0.128 eV, a difference between a-Si and Si was 0.256 eV, while between a-Si:H and pure Si a difference of 0.186 eV was found. If the broadening in the Si2p peak for bulk PS for our data is due to Si-H bonds, the FHWM would decrease with time as the sample oxidises. 

However, the fitted FWHM value is stable for all samples. Depth profile Ar sputtering of PS may however lead to a creation of amorphous Si (aSi) or amorphous hydrogenated Si (aSi:H). Ley et al. \cite{Ley:8} reported that these compounds have a 0.186 eV and 0.256 eV difference in binding energy compared to bulk crystalline Si. This broadening of the Si-2p spectra is comparable to what is observed for the Si-2p peaks for bulk PS in our samples.

\subsection{Composition of the Porous Silicon before and after ageing}

The elemental composition of the fitted Si-2p peaks for bulk and surface PS are presented in Table \ref{table:bulk} and Table \ref{table:surface}, respectively. The composition of surface PS, after 0 days, contains pure Si (Si$^0$), Si$_2$O (Si$^+$) and Si$_3$SiH. The composition between day 0 and day 1 for surface PS changes to some extent, where the amount of Si$^0$ decreases as Si$_3$SiH increases. However, between day 0 and day 1 for bulk PS, the composition of pure Si and Si$_3$SiH only changes minimally. This indicates that during oxidation, the surface of PS will first react with oxygen, capping the PS layer, leaving "bulk PS" unchanged during the first two days. During this natural oxidation process, some of the passivating Si-H surface bonds of freshly etched PS are replaced by Si-O bonds. After 7 days, the surface PS also contain the elements such as SiO, Si$_2$O$_3$, and SiO$_2$. The dominating oxidation state of Si up to this point has been Si$_2$O$_3$ (Si$^{3+}$). After 21 days, more oxidation of pure Si occurs in surface PS on the expense of pure Si, and the dominating oxidation state is SiO$_2$ (Si$^{4+}$).

\begin{table}[h!]
 \caption{The atomic percentages ($\pm$ 0.4 at. \%) of the compounds in bulk PS before and after oxidation.}	
\begin{tabular}{lllllll}
\hline
Days & Si$^0$ & Si$^{+}$ & Si$^{2+}$ & Si$^{3+}$ & Si$^{4+}$ & Si$_3$SiH \\
 \hline
 0 & 81.9 & 6.0 & 0 & 0 & 0 & 12.1 \\ 
 1 & 84.2 & 5.3 & 0 & 0 & 0 & 10.5 \\ 
 7 & 67.2 & 13.0 & 4.4 & 2.8 & 0  & 12.6 \\ 
 21 & 50.6 & 16.6 & 9.1 & 9.9 & 1.5 & 12.4 \\ 
 42 & 25.8 & 10.4 & 11.3 & 23.4 & 27.6 & 1.6 \\
\hline 
\end{tabular}
\label{table:bulk}
\end{table}

\begin{table}[h!]
 \caption{The atomic percentages ($\pm$ 0.4 at. \%) of the compounds in surface PS before and after oxidation.}	
\begin{tabular}{lllllll}
\hline
Days & Si$^0$ & Si$^{+}$ & Si$^{2+}$ & Si$^{3+}$ & Si$^{4+}$ & Si$_3$SiH  \\
 \hline
 0 & 43.8 & 12.8 & 0 & 0 & 0 & 43.4 \\ 
 1 & 50.2 & 12.5 & 0 & 0 & 0 & 37.3 \\ 
 7 & 37.4 & 14.5 & 4.5 & 8.7 & 1.2 & 33.6  \\ 
 21 & 28.1 & 10.5 & 8.3 & 20.6 & 11.1 & 21.4 \\ 
 42 & 14.3 & 7.4 & 5.1 & 12.4 & 45.3  & 15.4  \\
\hline 
\end{tabular}
\label{table:surface}
\end{table}

\subsection{Growth of SiO$_x$}

\begin{figure}  
	\begin{center}
    \includegraphics[width=0.5\textwidth]{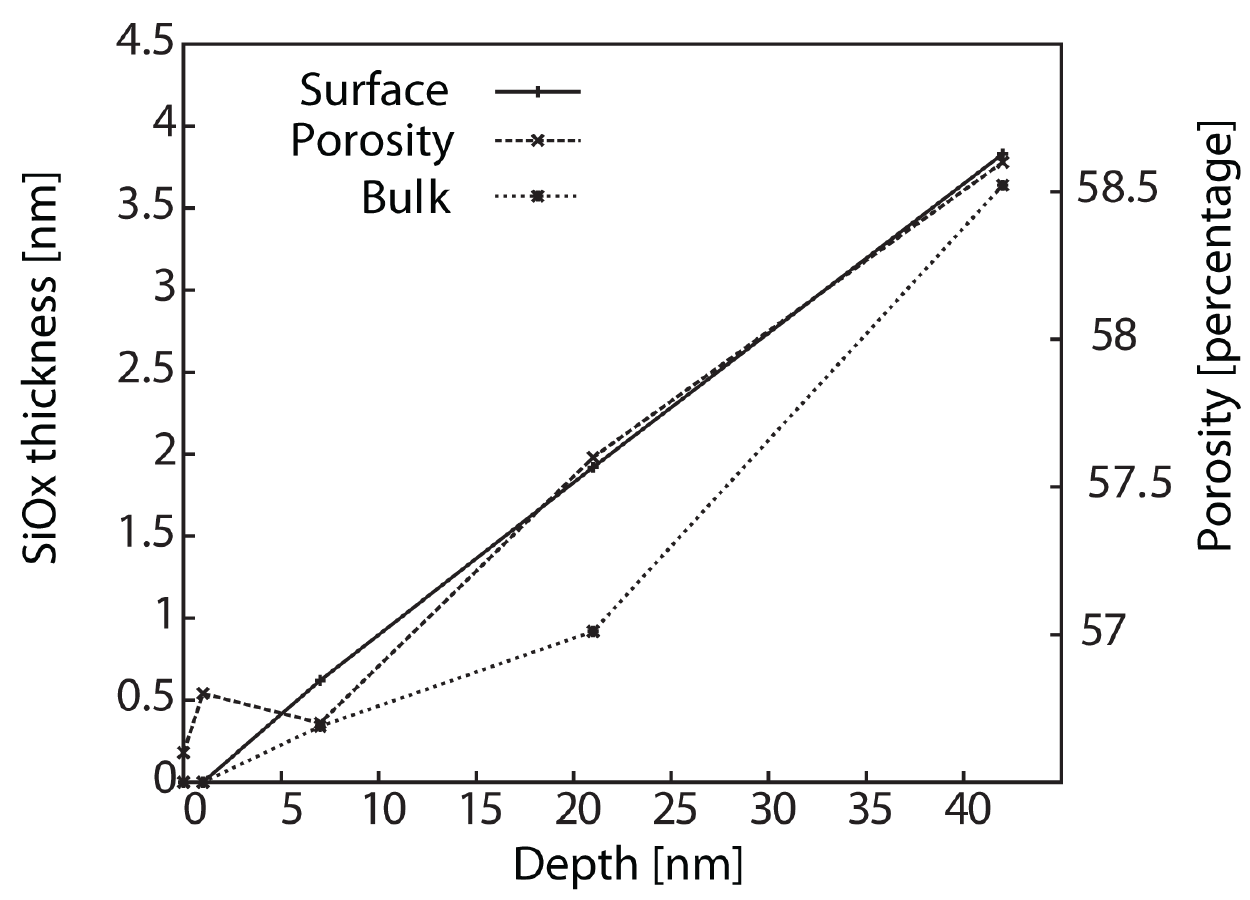}
   \caption{ The calculated SiO$_x$ thickness from XPS data, plotted with depth for the surface and bulk PS. In addition, the porosity of the PS with ageing, as modelled by ellipsometry, has also been added to the plot.}
    \label{figure:5}
  \end{center}
\end{figure}

The thickness of the SiO$_x$ layers of surface and bulk PS have been calculated by using the method described by Watts and Wolstenholme \cite{bok:watts}

\begin{equation}
d_{oxide} = \lambda_{SiO_2} \cos \Theta ln[1+ (R^{expt} /R^\infty)]
\end{equation}

where the mean free path for the Si 2p photoelectrons in SiO$_2$ ($\lambda_{SiO_2}$) is 3.7 nm,  the angle of emission$\Theta$ is 0, the ratio $R^{expt} = I^{exp}_{SiO_2}/I^{exp}_{Si}$, and the ratio $R^\infty$ is \cite{bok:watts}

\begin{equation}
R^\infty = \frac{\sigma_{Si,SiO_2} \lambda_{Si,SiO_2}}{\sigma_{Si,Si} \lambda_{Si,Si}}
\end{equation}

where $\sigma_{Si,SiO_2}$ is the atomic number density (atoms pr unit volume) of Si in SiO$_2$. 

\begin{equation}
\frac{\sigma_{Si,SiO_2}}{\sigma_{Si,Si}} = \frac{D_{SiO_2} F_{Si}}{D_{Si} F_{SiO_2}}
\end{equation}

where $D_{SiO_2}$ is the density (mass pr unit volume) of SiO$_2$ and $F_{Si}$ is the formula weight of Si \cite{bok:watts}. From these equations we calculated the $R^\infty$ to be 0.61. The thickness of the SiO$_x$ was then found for the ten samples, and plotted in Figure \ref{figure:5}. The plot shows that there is almost a linear relation of SiO$_x$ growth on both surface and bulk PS. The SiO$_x$ thickness at the surface PS grows from 0 nm to 3.8 nm during the 42 days. SiO$_x$ in the pores of bulk PS however, grows from 0 nm to 3.6 nm. This calculations is based on an oxide on a planer surface. Therefore, with a rough surface, such as for PS, an overestimation of the oxide will be carried out when calculating the SiO$_x$ thickness. However, this method of analysis does not take the sub-oxides into account, which in turn will result in a smaller thickness then what is expected. 

The calculated oxide thickness is far smaller then what was observed with TEM in Figure \ref{figure:2}, which was about 15 nm. If the oxide thickness after ageing was indeed 15 nm, no Si$^0$ peak would be visible in the spectra from this sample. This implies that the SiO$_x$ coverage is not uniformly distributed on the PS. The oxide thickness calculated in this paper is therefore the average oxide thickness.

\subsection{Refractive index and porosity of the Porous Silicon before and after oxidation.}

\begin{figure}  
	\begin{center}
    \includegraphics[width=0.5\textwidth]{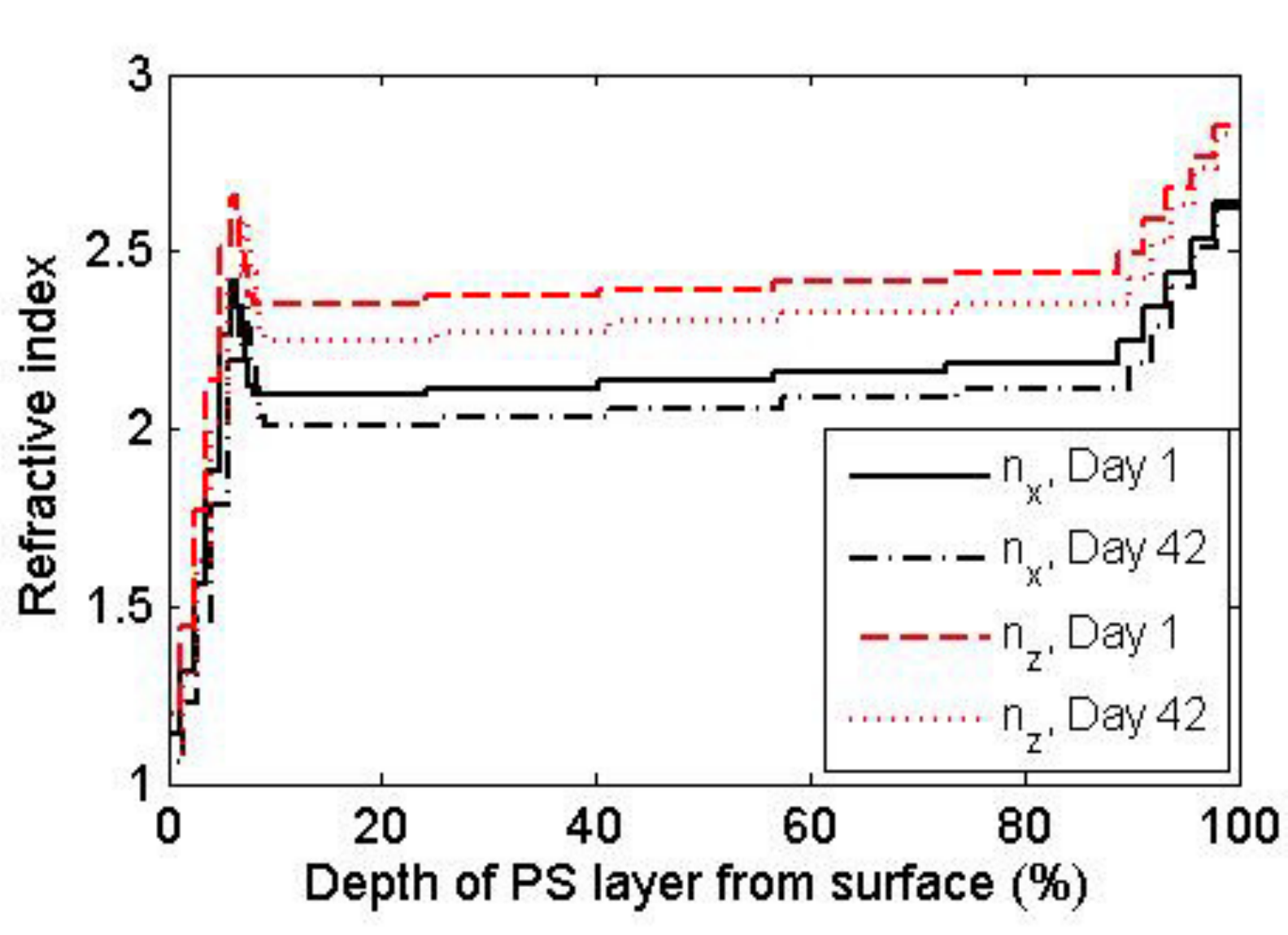}
   \caption{Change in refractive indices from day 1 to day 42.}
    \label{figure:6}
  \end{center}
\end{figure}

Figure \ref{figure:6} shows the refractive index with depth in the PS layer at day 1 and day 42. The layers are birefringent, so the refractive indices in both x- and z- (normal to the sample surface) directions are shown. As there is only one axis of anisotropy, the refractive index in the y-direction is identical to the x-direction and is not shown. In both directions, a small reduction in the refractive index is observed from day 1 to day 42.

For best possible accordance between SE and XPS, each sample is measured by ellipsometry within day one after etching and then again just before XPS measurements. Comparing the information obtained about oxidation of the PS structures, it is clear that SE is rather insensitive to oxidation in the PS structures. Partly, the inclusion of oxide in the ellipsometry model is complicated by the number of different oxidation states; for most of the samples only a small fraction of the oxidized species is SiO$_2$. It is also possible that an alternative model could give larger deflections with respect to oxygen content. However, when this is said, the marginal difference in $\Psi$ and $\Delta$, and consequently in refractive index between day 1 and day 42, indicates that it is a challenge to determine oxidation of PS with SE. It is also rather common to neglect oxidation when performing ellipsometric characterization of PS \cite{Selj:1, W:ny}. Whether neglecting SiO$_2$ is a viable approach or not depends, of course on the amount of oxide, but also on the information sought. Obviously, information about the chemical composition of the material is lost, but, on the other hand, changes in the effective refractive index of the material are small. Considering the densities and molecular weights of Si ($d_{Si}$ = 2330 kg/m3, $Mw_{Si}$ = 28.0855 g/mol) and SiO$_2$ ($d_{SiO_2}$ = 2200 kg/m3, $Mw_{SiO_2}$ = 60.0843 g/mol) it can be seen that for an oxide of thickness t, the fractional consumption of the Si wall is

\begin{equation}
(Mw_{Si}/d_{Si})/(Mw_{SiO_2}/d_{SiO_2}) = 0.44t
\end{equation}

while the remaining 0.56t is expansion into the pores (26). A locally flat surface, i.e. isotropic expansion through the volume (27), and oxidation by SiO$_2$ only is assumed. The net effect of oxidation can then be thought of as a replacement of a medium consisting of 44 \% Si and 56 \% air with a medium consisting of 100 \% SiO$_2$. In the Bruggeman effective medium model (BEMA), 44 \% Si and 56 \% air have an effective refractive index, N$_{eff}$, given by the equation

\begin{equation}
0.44 \frac{(N_{\text{Si}}^2 - N_{\text{eff}}^2)}{(N_{\text{Si}}^2 + 2N_{\text{eff}}^2)} + 0.56 \frac{(N_{\text{Air}}^2 - N_{\text{eff}}^2)}{(N_{\text{Air}}^2 + 2N_{\text{eff}}^2)} = 0
\end{equation}

Inserting the refractive indices N$_{Si}$(700 nm) = 3.77 and N$_{Air}$(700 nm) = 1.00, gives an effective refractive index N$_{eff}$ (700 nm) = 2.07. The refractive index of SiO$_2$ at the same wavelength is 1.45, hence the change in effective refractive index of the oxidized area is relatively small. The small reduction in refractive index gives the observed outcome of increased porosity in the ellipsometric modelling. The apparent increase in porosity with ageing is therefore ascribed to an increase in oxide content of the structure. As stated earlier, the oxidation growth is almost linear, this seems also to be the case for the porosity of the PS. The apparent increase in porosity is relatively modest and generates only a small change in the reflectance of these samples. However, adding SiO$_2$ in the ellipsometric modelling does not improve the fit for any of the samples.

\begin{figure}  
	\begin{center}
    \includegraphics[width=0.5\textwidth]{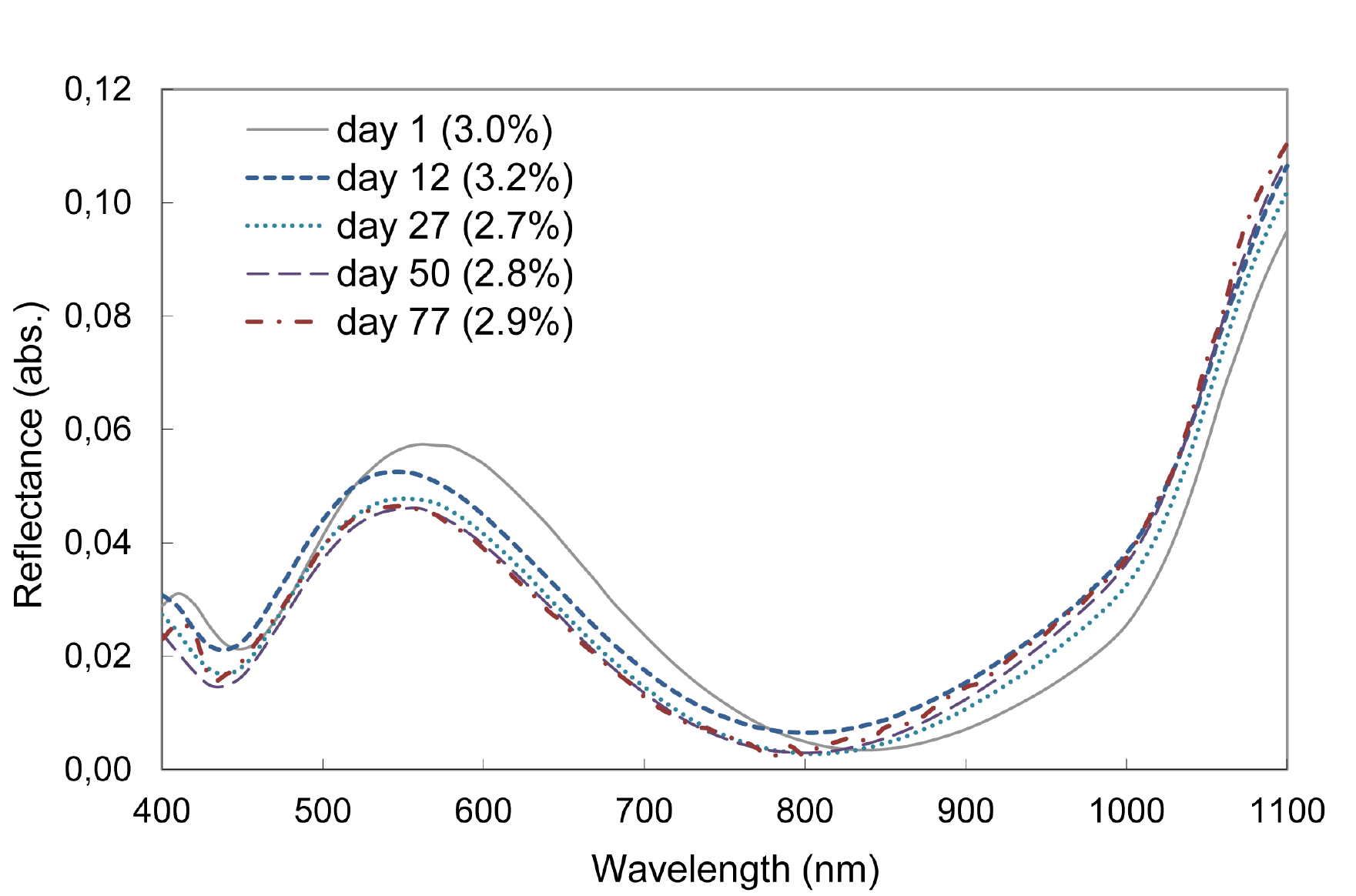}
   \caption{Evolution of the reflection from a PS ARC with time. The integrated reflectance is given in parenthesis.}
    \label{figure:7}
  \end{center}
\end{figure}

The evolution of the reflectance from a graded PS ARC with ageing is showed in Figure \ref{figure:7}. The reflectance integrated over the solar spectrum is shown in parenthesis. Reflectance measurements are performed immediately after fabrication and then with intervals of approximately two weeks. A small wavelength shift and amplitude lowering of the reflectance peak is visible. The uncertainty of the measurements is estimated to be $\sim$ 0.15 \% absolute, while the measured effective reflectance of the ARC varies by 0.5 \% over time. Therefore, it seems that the ARC experiences small variations in the reflectance with time, but the performance is more inclined towards an improvement than a degradation.

\section{Conclusion}

During oxidation of the PS elements such as pure Si (Si$^0$), Si$_2$O (Si$^+$), SiO (Si$^{2+}$), Si$_2$O$_3$ (Si$^{3+}$), and SiO$_2$ (Si$^{4+}$) is present. In addition, both hydrogen and carbon is introduced to the PS in the form of Si$_3$SiH and CO. Also, when sputtering the PS with Ar for depth profiling, aSi is created in the structure. This results in an increased FWHM of the Si2p peaks. Ageing the PS results in a linear increase in the average SiO$_x$ thickness, which grows from 0 nm to 3.8 nm during the 42 days. Bulk PS however, grows from 0 nm to 3.6 nm. The porosity shows the same trend, with a linear increase in porosity from 56.6 \% to 58.6 \% with ageing and the increase in SiO$_x$ thickness. The reflectance of the multilayer structures is apparently quite robust and not subject to any degradation. If structure and storage conditions are known, the small change in reflectance due to oxidation could be accounted for in the design of the coatings to improve the effective reflectivity further.


\newpage
\newpage

\begin{thebibliography}{31}
\expandafter\ifx\csname natexlab\endcsname\relax\def\natexlab#1{#1}\fi
\expandafter\ifx\csname bibnamefont\endcsname\relax
  \def\bibnamefont#1{#1}\fi
\expandafter\ifx\csname bibfnamefont\endcsname\relax
  \def\bibfnamefont#1{#1}\fi
\expandafter\ifx\csname citenamefont\endcsname\relax
  \def\citenamefont#1{#1}\fi
\expandafter\ifx\csname url\endcsname\relax
  \def\url#1{\texttt{#1}}\fi
\expandafter\ifx\csname urlprefix\endcsname\relax\def\urlprefix{URL }\fi
\providecommand{\bibinfo}[2]{#2}
\providecommand{\eprint}[2][]{\url{#2}}

\bibitem[{\citenamefont{Zhang}(2004)}]{Zhang:ny}
\bibinfo{author}{\bibfnamefont{X.~G.} \bibnamefont{Zhang}},
  \bibinfo{journal}{Journal of The Electrochemical Society}
  \textbf{\bibinfo{volume}{151}}, \bibinfo{pages}{C69} (\bibinfo{year}{2004}).

\bibitem[{\citenamefont{Rossow et~al.}(1995{\natexlab{a}})\citenamefont{Rossow,
  Frotscher, Thonissen, Berger, and Frohnhoff}}]{Rossow:ny}
\bibinfo{author}{\bibfnamefont{U.}~\bibnamefont{Rossow}},
  \bibinfo{author}{\bibfnamefont{U.}~\bibnamefont{Frotscher}},
  \bibinfo{author}{\bibfnamefont{M.}~\bibnamefont{Thonissen}},
  \bibinfo{author}{\bibfnamefont{M.~G.} \bibnamefont{Berger}},
  \bibnamefont{and}
  \bibinfo{author}{\bibfnamefont{S.}~\bibnamefont{Frohnhoff}},
  \bibinfo{journal}{Thin Solid Films} \textbf{\bibinfo{volume}{255}},
  \bibinfo{pages}{5} (\bibinfo{year}{1995}{\natexlab{a}}).

\bibitem[{\citenamefont{Yuan et~al.}(2009)\citenamefont{Yuan, Yost, Page,
  Stradins, Meier, and Branz}}]{Yuan:ny}
\bibinfo{author}{\bibfnamefont{H.}~\bibnamefont{Yuan}},
  \bibinfo{author}{\bibfnamefont{V.~E.} \bibnamefont{Yost}},
  \bibinfo{author}{\bibfnamefont{M.~R.} \bibnamefont{Page}},
  \bibinfo{author}{\bibfnamefont{P.}~\bibnamefont{Stradins}},
  \bibinfo{author}{\bibfnamefont{D.~L.} \bibnamefont{Meier}}, \bibnamefont{and}
  \bibinfo{author}{\bibfnamefont{H.}~\bibnamefont{Branz}},
  \bibinfo{journal}{Appl. Phys. Lett.} \textbf{\bibinfo{volume}{95}},
  \bibinfo{pages}{123501} (\bibinfo{year}{2009}).

\bibitem[{\citenamefont{Kwon et~al.}(2007)\citenamefont{Kwon, Lee, and
  Ju}}]{Kwon:ny}
\bibinfo{author}{\bibfnamefont{J.}~\bibnamefont{Kwon}},
  \bibinfo{author}{\bibfnamefont{S.}~\bibnamefont{Lee}}, \bibnamefont{and}
  \bibinfo{author}{\bibfnamefont{B.}~\bibnamefont{Ju}}, \bibinfo{journal}{J.
  Appl. Phys.} \textbf{\bibinfo{volume}{101}}, \bibinfo{pages}{104515}
  (\bibinfo{year}{2007}).

\bibitem[{\citenamefont{Lipinski et~al.}(2003)\citenamefont{Lipinski, Bastide,
  Panek, and Levy-Clement}}]{Lipinski:ny}
\bibinfo{author}{\bibfnamefont{M.}~\bibnamefont{Lipinski}},
  \bibinfo{author}{\bibfnamefont{S.}~\bibnamefont{Bastide}},
  \bibinfo{author}{\bibfnamefont{P.}~\bibnamefont{Panek}}, \bibnamefont{and}
  \bibinfo{author}{\bibfnamefont{C.}~\bibnamefont{Levy-Clement}},
  \bibinfo{journal}{Physica status solidi(a)} \textbf{\bibinfo{volume}{197}},
  \bibinfo{pages}{512} (\bibinfo{year}{2003}).

\bibitem[{\citenamefont{Yerokhov et~al.}(2000)\citenamefont{Yerokhov, Melnyk,
  Tsisaruk, and Semochko}}]{Yerokhov:ny}
\bibinfo{author}{\bibfnamefont{V.}~\bibnamefont{Yerokhov}},
  \bibinfo{author}{\bibfnamefont{I.}~\bibnamefont{Melnyk}},
  \bibinfo{author}{\bibfnamefont{A.}~\bibnamefont{Tsisaruk}}, \bibnamefont{and}
  \bibinfo{author}{\bibfnamefont{I.}~\bibnamefont{Semochko}},
  \bibinfo{journal}{Opto-elect. rev.} \textbf{\bibinfo{volume}{8}},
  \bibinfo{pages}{414} (\bibinfo{year}{2000}).

\bibitem[{\citenamefont{Bilyalov et~al.}(2000)\citenamefont{Bilyalov, Ludemann,
  Wettling, Staalmans, Poortmans, Nijs, Schirone, Sotgiu, S., and
  Levy-Clemet}}]{Wettling:ny}
\bibinfo{author}{\bibfnamefont{R.}~\bibnamefont{Bilyalov}},
  \bibinfo{author}{\bibfnamefont{R.}~\bibnamefont{Ludemann}},
  \bibinfo{author}{\bibfnamefont{W.}~\bibnamefont{Wettling}},
  \bibinfo{author}{\bibfnamefont{L.}~\bibnamefont{Staalmans}},
  \bibinfo{author}{\bibfnamefont{J.}~\bibnamefont{Poortmans}},
  \bibinfo{author}{\bibfnamefont{J.}~\bibnamefont{Nijs}},
  \bibinfo{author}{\bibfnamefont{L.}~\bibnamefont{Schirone}},
  \bibinfo{author}{\bibfnamefont{G.}~\bibnamefont{Sotgiu}},
  \bibinfo{author}{\bibfnamefont{S.}~\bibnamefont{S.}}, \bibnamefont{and}
  \bibinfo{author}{\bibfnamefont{C.}~\bibnamefont{Levy-Clemet}},
  \bibinfo{journal}{Solar Energy Materials and Solar Cells}
  \textbf{\bibinfo{volume}{60}}, \bibinfo{pages}{391} (\bibinfo{year}{2000}).

\bibitem[{\citenamefont{Selj et~al.}(2010)\citenamefont{Selj, Th{\o }gersen,
  Foss, and Marstein}}]{selj:2}
\bibinfo{author}{\bibfnamefont{J.~H.} \bibnamefont{Selj}},
  \bibinfo{author}{\bibfnamefont{A.}~\bibnamefont{Th{\o }gersen}},
  \bibinfo{author}{\bibfnamefont{S.}~\bibnamefont{Foss}}, \bibnamefont{and}
  \bibinfo{author}{\bibfnamefont{E.}~\bibnamefont{Marstein}},
  \bibinfo{journal}{J. Appl Phys.} \textbf{\bibinfo{volume}{107}},
  \bibinfo{pages}{074904} (\bibinfo{year}{2010}).

\bibitem[{\citenamefont{Herino et~al.}(1987)\citenamefont{Herino, Bomchil,
  Barla, Bertrand, and Ginoux}}]{Herino:PS}
\bibinfo{author}{\bibfnamefont{R.}~\bibnamefont{Herino}},
  \bibinfo{author}{\bibfnamefont{G.}~\bibnamefont{Bomchil}},
  \bibinfo{author}{\bibfnamefont{K.}~\bibnamefont{Barla}},
  \bibinfo{author}{\bibfnamefont{C.}~\bibnamefont{Bertrand}}, \bibnamefont{and}
  \bibinfo{author}{\bibfnamefont{J.~L.} \bibnamefont{Ginoux}},
  \bibinfo{journal}{J. Electrochem. Soc.} \textbf{\bibinfo{volume}{134}}
  (\bibinfo{year}{1987}).

\bibitem[{\citenamefont{Canham}(1997)}]{Canham:1}
\bibinfo{author}{\bibfnamefont{L.}~\bibnamefont{Canham}},
  \emph{\bibinfo{title}{Properties of Porous Silicon}}
  (\bibinfo{publisher}{EMIS Datareviews Series No 18}, \bibinfo{year}{1997}).

\bibitem[{\citenamefont{Grosman and Ortega}(2006{\natexlab{a}})}]{Grosman:1}
\bibinfo{author}{\bibfnamefont{A.}~\bibnamefont{Grosman}} \bibnamefont{and}
  \bibinfo{author}{\bibfnamefont{C.}~\bibnamefont{Ortega}},
  \emph{\bibinfo{title}{Properties of Porous Silicon EMIS Data Reviews Series
  No 18.}} (\bibinfo{publisher}{The institution of Electrical Engineers,
  London, UK}, \bibinfo{year}{2006}{\natexlab{a}}).

\bibitem[{\citenamefont{Grosman and Ortega}(2006{\natexlab{b}})}]{grosman:16}
\bibinfo{author}{\bibfnamefont{A.}~\bibnamefont{Grosman}} \bibnamefont{and}
  \bibinfo{author}{\bibfnamefont{C.}~\bibnamefont{Ortega}},
  \emph{\bibinfo{title}{Properties of Porous Silicon.}}
  (\bibinfo{publisher}{The institution of Electrical Engineers, London, UK},
  \bibinfo{year}{2006}{\natexlab{b}}), p. \bibinfo{pages}{145}.

\bibitem[{\citenamefont{Rao et~al.}(1990)\citenamefont{Rao, Ozanam, and
  Chazalviel}}]{Rao:19}
\bibinfo{author}{\bibfnamefont{A.}~\bibnamefont{Rao}},
  \bibinfo{author}{\bibfnamefont{F.}~\bibnamefont{Ozanam}}, \bibnamefont{and}
  \bibinfo{author}{\bibfnamefont{J.-N.} \bibnamefont{Chazalviel}},
  \bibinfo{journal}{J. of Elec. Spec. and Rel. Phenom.}
  \textbf{\bibinfo{volume}{54-55}}, \bibinfo{pages}{1215}
  (\bibinfo{year}{1990}).

\bibitem[{\citenamefont{Domashevskaya et~al.}(1998)\citenamefont{Domashevskaya,
  Kashkarov, Manukovskii, Schukarev, and Terekhov}}]{Domashevskaya:13}
\bibinfo{author}{\bibfnamefont{E.}~\bibnamefont{Domashevskaya}},
  \bibinfo{author}{\bibfnamefont{V.}~\bibnamefont{Kashkarov}},
  \bibinfo{author}{\bibfnamefont{E.}~\bibnamefont{Manukovskii}},
  \bibinfo{author}{\bibfnamefont{A.}~\bibnamefont{Schukarev}},
  \bibnamefont{and} \bibinfo{author}{\bibfnamefont{V.}~\bibnamefont{Terekhov}},
  \bibinfo{journal}{J. of. Elec. Spec. Rel. Phenom.}
  \textbf{\bibinfo{volume}{88}}, \bibinfo{pages}{969}
  (\bibinfo{year}{1998}).

\bibitem[{\citenamefont{Peden et~al.}(1993)\citenamefont{Peden, Rogers, Jr.,
  Kidd, and Tsang.}}]{Peden:1}
\bibinfo{author}{\bibfnamefont{C.~H.~F.} \bibnamefont{Peden}},
  \bibinfo{author}{\bibfnamefont{J.~W.} \bibnamefont{Rogers}},
  \bibinfo{author}{\bibfnamefont{N.~D.~S.} \bibnamefont{Jr.}},
  \bibinfo{author}{\bibfnamefont{K.~B.} \bibnamefont{Kidd}}, \bibnamefont{and}
  \bibinfo{author}{\bibfnamefont{K.~L.} \bibnamefont{Tsang.}},
  \bibinfo{journal}{Phys. Rev. B} \textbf{\bibinfo{volume}{47}},
  \bibinfo{pages}{15622} (\bibinfo{year}{1993}).

\bibitem[{\citenamefont{Himpsel et~al.}(1988)\citenamefont{Himpsel, McFeeley,
  Taleb-Ibrahimi, Yarmoff, and Hollinger.}}]{Himpsel:2}
\bibinfo{author}{\bibfnamefont{F.}~\bibnamefont{Himpsel}},
  \bibinfo{author}{\bibfnamefont{F.}~\bibnamefont{McFeeley}},
  \bibinfo{author}{\bibfnamefont{A.}~\bibnamefont{Taleb-Ibrahimi}},
  \bibinfo{author}{\bibfnamefont{J.}~\bibnamefont{Yarmoff}}, \bibnamefont{and}
  \bibinfo{author}{\bibfnamefont{G.}~\bibnamefont{Hollinger.}},
  \bibinfo{journal}{Phys. Rev. B} \textbf{\bibinfo{volume}{38}},
  \bibinfo{pages}{6084} (\bibinfo{year}{1988}).

\bibitem[{\citenamefont{Th{\o }gersen et~al.}(2008)\citenamefont{Th{\o }gersen,
  Diplas, Mayandi, Finstad, Olsen, Watts, Mitome, and Bando}}]{Thøgersen:3}
\bibinfo{author}{\bibfnamefont{A.}~\bibnamefont{Th{\o }gersen}},
  \bibinfo{author}{\bibfnamefont{S.}~\bibnamefont{Diplas}},
  \bibinfo{author}{\bibfnamefont{J.}~\bibnamefont{Mayandi}},
  \bibinfo{author}{\bibfnamefont{T.}~\bibnamefont{Finstad}},
  \bibinfo{author}{\bibfnamefont{A.}~\bibnamefont{Olsen}},
  \bibinfo{author}{\bibfnamefont{J.~F.} \bibnamefont{Watts}},
  \bibinfo{author}{\bibfnamefont{M.}~\bibnamefont{Mitome}}, \bibnamefont{and}
  \bibinfo{author}{\bibfnamefont{Y.}~\bibnamefont{Bando}}, \bibinfo{journal}{J.
  of Appl. Phys.} \textbf{\bibinfo{volume}{103}}, \bibinfo{pages}{024308}
  (\bibinfo{year}{2008}).

\bibitem[{\citenamefont{Selj et~al.}(2011)\citenamefont{Selj, Marstein, Th{\o
  }gersen, and Foss}}]{Selj:1}
\bibinfo{author}{\bibfnamefont{J.}~\bibnamefont{Selj}},
  \bibinfo{author}{\bibfnamefont{E.}~\bibnamefont{Marstein}},
  \bibinfo{author}{\bibfnamefont{A.}~\bibnamefont{Th{\o }gersen}},
  \bibnamefont{and} \bibinfo{author}{\bibfnamefont{S.}~\bibnamefont{Foss}},
  \bibinfo{journal}{Status Physica Solidi (c)} \textbf{\bibinfo{volume}{8}},
  \bibinfo{pages}{1860} (\bibinfo{year}{2011}).

\bibitem[{\citenamefont{http://www.casaxps.com}()}]{casa:xps}
\bibinfo{author}{\bibnamefont{http://www.casaxps.com}}.

\bibitem[{\citenamefont{Pettersson et~al.}(1998)\citenamefont{Pettersson,
  Hultman, and Arwin}}]{Pettersson:JHS}
\bibinfo{author}{\bibfnamefont{L.~A.~A.} \bibnamefont{Pettersson}},
  \bibinfo{author}{\bibfnamefont{L.}~\bibnamefont{Hultman}}, \bibnamefont{and}
  \bibinfo{author}{\bibfnamefont{H.}~\bibnamefont{Arwin}},
  \bibinfo{journal}{Appl. Opt.} \textbf{\bibinfo{volume}{37}},
  \bibinfo{pages}{4130} (\bibinfo{year}{1998}).

\bibitem[{\citenamefont{Rossow et~al.}(1995{\natexlab{b}})\citenamefont{Rossow,
  Frotscher, Thonissen, Berger, Frohnhoff, Munder, and Richter}}]{Rossow:JHS}
\bibinfo{author}{\bibfnamefont{U.}~\bibnamefont{Rossow}},
  \bibinfo{author}{\bibfnamefont{U.}~\bibnamefont{Frotscher}},
  \bibinfo{author}{\bibfnamefont{M.}~\bibnamefont{Thonissen}},
  \bibinfo{author}{\bibfnamefont{M.~G.} \bibnamefont{Berger}},
  \bibinfo{author}{\bibfnamefont{S.}~\bibnamefont{Frohnhoff}},
  \bibinfo{author}{\bibfnamefont{H.}~\bibnamefont{Munder}}, \bibnamefont{and}
  \bibinfo{author}{\bibfnamefont{W.}~\bibnamefont{Richter}},
  \bibinfo{journal}{Thin Solid Films} \textbf{\bibinfo{volume}{255}},
  \bibinfo{pages}{5} (\bibinfo{year}{1995}{\natexlab{b}}).

\bibitem[{\citenamefont{Cerofolini et~al.}(2003)\citenamefont{Cerofolini,
  Galati, and Renna}}]{Cerofolini:6}
\bibinfo{author}{\bibfnamefont{G.~F.} \bibnamefont{Cerofolini}},
  \bibinfo{author}{\bibfnamefont{C.}~\bibnamefont{Galati}}, \bibnamefont{and}
  \bibinfo{author}{\bibfnamefont{L.}~\bibnamefont{Renna}},
  \bibinfo{journal}{Surf. Interface Anal.} \textbf{\bibinfo{volume}{35}},
  \bibinfo{pages}{968} (\bibinfo{year}{2003}).

\bibitem[{\citenamefont{Lu et~al.}(1993)\citenamefont{Lu, Graham, Jiang, and
  Tan}}]{Lu:18}
\bibinfo{author}{\bibfnamefont{Z.~H.} \bibnamefont{Lu}},
  \bibinfo{author}{\bibfnamefont{M.}~\bibnamefont{Graham}},
  \bibinfo{author}{\bibfnamefont{D.}~\bibnamefont{Jiang}}, \bibnamefont{and}
  \bibinfo{author}{\bibfnamefont{K.}~\bibnamefont{Tan}},
  \bibinfo{journal}{Appl. Phys. Lett.} \textbf{\bibinfo{volume}{63}},
  \bibinfo{pages}{2941} (\bibinfo{year}{1993}).

\bibitem[{\citenamefont{Pleul et~al.}(2003)\citenamefont{Pleul, Frenzel,
  Eschner, and Simon}}]{Eschner:17}
\bibinfo{author}{\bibfnamefont{D.}~\bibnamefont{Pleul}},
  \bibinfo{author}{\bibfnamefont{R.}~\bibnamefont{Frenzel}},
  \bibinfo{author}{\bibfnamefont{M.}~\bibnamefont{Eschner}}, \bibnamefont{and}
  \bibinfo{author}{\bibfnamefont{F.}~\bibnamefont{Simon}},
  \bibinfo{journal}{Anal Bioanal Chem} \textbf{\bibinfo{volume}{375}},
  \bibinfo{pages}{1276} (\bibinfo{year}{2003}).

\bibitem[{\citenamefont{Wagner et~al.}(1982)\citenamefont{Wagner, Passoja,
  Hillery, Kinisky, Six, Jansen, and Taylor}}]{Wagner:5}
\bibinfo{author}{\bibfnamefont{C.~D.} \bibnamefont{Wagner}},
  \bibinfo{author}{\bibfnamefont{D.~E.} \bibnamefont{Passoja}},
  \bibinfo{author}{\bibfnamefont{H.~F.} \bibnamefont{Hillery}},
  \bibinfo{author}{\bibfnamefont{T.~G.} \bibnamefont{Kinisky}},
  \bibinfo{author}{\bibfnamefont{H.~A.} \bibnamefont{Six}},
  \bibinfo{author}{\bibfnamefont{W.~T.} \bibnamefont{Jansen}},
  \bibnamefont{and} \bibinfo{author}{\bibfnamefont{J.~A.}
  \bibnamefont{Taylor}}, \bibinfo{journal}{J. Vac. Sci. Technol. A}
  \textbf{\bibinfo{volume}{21}}, \bibinfo{pages}{933} (\bibinfo{year}{1982}).

\bibitem[{\citenamefont{Hollinger et~al.}(1975)\citenamefont{Hollinger, Jugnet,
  Pertosa, and Duc}}]{Hollinger:SiO2}
\bibinfo{author}{\bibfnamefont{G.}~\bibnamefont{Hollinger}},
  \bibinfo{author}{\bibfnamefont{Y.}~\bibnamefont{Jugnet}},
  \bibinfo{author}{\bibfnamefont{P.}~\bibnamefont{Pertosa}}, \bibnamefont{and}
  \bibinfo{author}{\bibfnamefont{T.~M.} \bibnamefont{Duc}},
  \bibinfo{journal}{Chem. Phys. Lett.} \textbf{\bibinfo{volume}{36}},
  \bibinfo{pages}{441} (\bibinfo{year}{1975}).

\bibitem[{\citenamefont{Yue et~al.}(1999)\citenamefont{Yue, Jiang, Wang,
  Gardner, and Jr.}}]{yue:xps}
\bibinfo{author}{\bibfnamefont{Z.}~\bibnamefont{Yue}},
  \bibinfo{author}{\bibfnamefont{W.}~\bibnamefont{Jiang}},
  \bibinfo{author}{\bibfnamefont{L.}~\bibnamefont{Wang}},
  \bibinfo{author}{\bibfnamefont{S.}~\bibnamefont{Gardner}}, \bibnamefont{and}
  \bibinfo{author}{\bibfnamefont{C.~P.} \bibnamefont{Jr.}},
  \bibinfo{journal}{Carbon} \textbf{\bibinfo{volume}{37}},
  \bibinfo{pages}{1785} (\bibinfo{year}{1999}).

\bibitem[{\citenamefont{Guerrero-Lemus
  et~al.}(1999)\citenamefont{Guerrero-Lemus, Moreno, Martin-Palma, Ben-Hander,
  Martinez-Duart, Fierro, and Gomez-Garrido}}]{GuerreroLemus:7}
\bibinfo{author}{\bibfnamefont{R.}~\bibnamefont{Guerrero-Lemus}},
  \bibinfo{author}{\bibfnamefont{J.}~\bibnamefont{Moreno}},
  \bibinfo{author}{\bibfnamefont{R.}~\bibnamefont{Martin-Palma}},
  \bibinfo{author}{\bibfnamefont{F.}~\bibnamefont{Ben-Hander}},
  \bibinfo{author}{\bibfnamefont{J.~M.} \bibnamefont{Martinez-Duart}},
  \bibinfo{author}{\bibfnamefont{J.}~\bibnamefont{Fierro}}, \bibnamefont{and}
  \bibinfo{author}{\bibfnamefont{P.}~\bibnamefont{Gomez-Garrido}},
  \bibinfo{journal}{Thin Solid Films} \textbf{\bibinfo{volume}{354}},
  \bibinfo{pages}{34} (\bibinfo{year}{1999}).

\bibitem[{\citenamefont{Ley et~al.}(1982)\citenamefont{Ley, Reichardt, and
  Johnson}}]{Ley:8}
\bibinfo{author}{\bibfnamefont{L.}~\bibnamefont{Ley}},
  \bibinfo{author}{\bibfnamefont{J.}~\bibnamefont{Reichardt}},
  \bibnamefont{and} \bibinfo{author}{\bibfnamefont{R.~L.}
  \bibnamefont{Johnson}}, \bibinfo{journal}{Phys. Rev. Lett.}
  \textbf{\bibinfo{volume}{49}}, \bibinfo{pages}{1664} (\bibinfo{year}{1982}).

\bibitem[{\citenamefont{Watts and Wolstenholme}(2003)}]{bok:watts}
\bibinfo{author}{\bibfnamefont{J.}~\bibnamefont{Watts}} \bibnamefont{and}
  \bibinfo{author}{\bibfnamefont{J.}~\bibnamefont{Wolstenholme}},
  \emph{\bibinfo{title}{An introduction to Surface Analysis by XPS and AES}}
  (\bibinfo{publisher}{WILEY}, \bibinfo{year}{2003}).

\bibitem[{\citenamefont{Wongmanerod et~al.}(2001)\citenamefont{Wongmanerod,
  Zangooie, and Arwin}}]{W:ny}
\bibinfo{author}{\bibfnamefont{C.}~\bibnamefont{Wongmanerod}},
  \bibinfo{author}{\bibfnamefont{S.}~\bibnamefont{Zangooie}}, \bibnamefont{and}
  \bibinfo{author}{\bibfnamefont{H.}~\bibnamefont{Arwin}},
  \bibinfo{journal}{Appl. Surf. Sci.} \textbf{\bibinfo{volume}{172}},
  \bibinfo{pages}{117} (\bibinfo{year}{2001}).

\end{thebibliography}

\end{document}